\documentclass [preprint]{aastex}
\singlespace
\shorttitle{Mid-Infrared Spectroscopy of Luminous Infrared Galaxies}
\shortauthors{Soifer  et al.}

\begin{document}

\title{Mid-Infrared Spectroscopy of Infrared Luminous Galaxies with 
Sub-Arcsecond Resolution
\altaffilmark{1}}

\author{B. T. Soifer\altaffilmark{2}, G. Neugebauer\altaffilmark{3}, 
K. Matthews, E. Egami\altaffilmark{3}}
\affil{Caltech Optical Observatories, California Institute of
Technology, 320-47, Pasadena, CA 91125}
\email { bts@irastro.caltech.edu, gxn@irastro.caltech.edu, kym@caltech.edu, 
eegami@as.arizona.edu}
\author{A. J. Weinberger\altaffilmark{4}}
\affil{Department of Physics and Astronomy, University of California Los
Angeles, 156205 Los~Angeles, CA 90095}
\email {weinberger@dtm.ciw.edu}

\altaffiltext{1} {based on observations obtained at the
W. M. Keck Observatory  which is operated as a scientific partnership
among the  California Institute of Technology, the  University of
California and the National Aeronautics and Space Administration.}

\altaffiltext{2} { also at SIRTF Science Center, California Institute of
Technology, 220-6, Pasadena, CA 91125}

\altaffiltext{3} { present address: Steward Observatory, University of
Arizona,
Tucson, AZ 85721}
\altaffiltext{4} { present address: Carnegie Institution of Washington, DTM,
Washington, D.C. 20015}

\vfill\eject

\begin{abstract}
Low spectral resolution ($\Delta\lambda$/$\lambda\sim$50) mid-infrared 
observations with high angular resolution (0.3--0.5$''$) using the
Long Wavelength Spectrometer on the Keck I Telescope are reported
of the nuclei of five highly luminous infrared bright galaxies. Spectra 
of eight distinct nuclei, ranging in luminosity from $\sim 10^{11}$L$_
{\odot}$ to $>10^{12} $L$_{\odot}$ have been obtained.  Four of the nuclei 
show the characteristic PAH emission features, i.e. 11.3$\mu$m emission as
well as the 8.6$\mu$m shoulder of the 7.7$\mu$m band.  The other nuclei 
show either weak PAH emission bands or no evidence for these bands.

The high spatial resolution of the observations reveals extended emission
in the 11.3$\mu$m PAH band associated with several of the compact
nuclear sources.  When proper account is taken of the
diffuse PAH emission, most of the compact sources show little or no
directly associated PAH emission.
The diffuse PAH emission is extended over spatial scales of 100--500 pc;
its presence shows that there is significant circumnuclear
UV/optical emission exciting the aromatic bands, most likely associated
with circumnuclear starbursts.

After the spectra of the nuclear sources are corrected for the
spectrum of the diffuse PAH emission, the peak apparent silicate optical
depth at 9.7$\mu$m can be as great as 15, corresponding to $>$ 150
magnitudes of visible light extinction.  Because of the large
silicate optical
depths, mid-infrared spectra are not probing  the nature of the
true nuclei in the most opaque compact sources.

\end{abstract}


\newpage

\section{Introduction}

Submillimeter observations (e.g. Eales et al. 2000, Blain et al. 1999) 
suggest that ultraluminous infrared galaxies (ULIRGs) at high redshift 
can account for the cosmic IR background (cf. Hauser \& Dwek, 2001).  
If powered by star formation, these ULIRGs can account for as much star 
formation as the galaxies detected in the UV continuum.  Alternatively, 
if the infrared luminosity is attributed to dust-enshrouded AGN, the cosmic
IR background could be measuring the  integrated accretion power
generated over the history of the universe.

Such observations illustrate the importance of understanding the origin
of the luminosity in infrared bright galaxies. Because these galaxies
are by their nature dusty, the normal  optical/near infrared
diagnostics break down, since the luminosity sources are obscured by many
magnitudes of visual extinction.  Attributing the luminosity to
dust enshrouded star formation or accretion onto black holes has
drastically different consequences for assessing the star formation
history of the universe.

Recent spectroscopic observations from ISO  have suggested that
spectroscopic diagnostics of the underlying power sources do exist
within the  mid-infrared.  Genzel et al (1998) suggested that there is
a correspondence  between  the presence of strong aromatic features 
(Polycyclic Aromatic Hydrocarbons, commonly referred to as PAHs)
at 6.2, 7.7, 8.6 and 11.3 $\mu$m and other diagnostics of
stellar ionizing radiation from young, massive stars,  while weak or
non-existent aromatic features are associated with infrared galaxies
whose luminosity is generated via AGNs.

The ISO observations were hampered by the large observing beams, which
included nearly all the infrared light from the  galaxies, thereby
combining both nuclear and circumnuclear emission.  In Seyfert
galaxies such as NGC 1068 and NGC 7469 (cf. Myer \& Scoville 1987,
Wilson et al. 1991), large observing beams include both a nuclear AGN
and a circumstellar starburst, so that the integrated spectrum is a
combination of radiation generated in multiple regions with widely
varying excitation conditions.

Ground-based, airborne and space-based observations, e.g. Gillett 
et al. (1975a), Willner et al. (1977), Helou et al. (2000), Sturm 
et al.(2000) have demonstrated that the full suite of PAH features 
is readily seen in many galaxies. Indeed, as shown by  Genzel et al.
(1998), Lutz et al.(1998), Rigopoulou et al. (1999) and Tran et al. 
(2001) these bands can dominate the mid-infrared spectra of ULIRGs.  
The 11.3$\mu$m feature is readily detectable, having a
line to continuum strength roughly half that of the 7.7$\mu$m band in
nearby galaxies such as M82 (Gillett et al., Sturm et al.).
Thus, the detection of PAH emission
via the 11.3$\mu$m band should  be nearly equivalent to using
the broader 7.7$\mu$m complex of features.  The presence of the
9.7$\mu$m silicate feature, which can produce strong absorption over
the entire 8-13 $\mu$m region, complicates the 
interpretation of mid-infrared spectra. Silicate absorption
has been shown to be quite strong
in some galaxy spectra (Roche et al. 1991).

The upcoming launch of SIRTF, with its capability of studying the
spectra of ULIRGs to redshifts substantially greater than one,
highlights the need to develop infrared diagnostics of the luminosity
in dusty, luminous galaxies. We have obtained low spectral resolution
($\lambda$/$\Delta\lambda\sim$50) mid-infrared observations from 8 --
12.5 $\mu$m of a sample of infrared luminous galaxies with
0.3--0.5$''$ spatial resolution using the Long Wavelength Spectrometer
(LWS) on the Keck I Telescope. With this spatial resolution  we have
been able to isolate individual compact sources that are emitting
copiously in the infrared and are our best local examples of ULIRGs.

Spectroscopy of some of the sources observed here has been
reported by Smith, Aitken and Roche (1989), Roche et al. (1991) and
Dudley (1999).  The data reported here are obtained with substantially
higher spatial resolution that has allowed us to 
resolve the separate nuclei in the multiple nuclei sources as
well as to explicitly detect spatially extended PAH emission associated
with the nuclei.

\section {The Sample}

The objects observed were selected to be high luminosity  galaxies
that are bright at 12 $\mu$m.  VV~114, Markarian~273, IRAS~08572+3915,
and Arp~220 were all selected from the IRAS Bright Galaxy  Sample
(Soifer, et al. 1987), while Markarian~463E is included in the Ultraluminous
Warm Galaxy Sample (Sanders, et al. 1988a). The total infrared
luminosities of the  systems range from 4 $\times$ 10$^{11}$
L$_{\odot}$ for VV~114 ( Soifer et al.) to 1.5 $\times$  10$^{12}$
L$_{\odot}$ for Arp~220 (Sanders et al. 1988b).  Mid-infrared imaging
with resolution  of 0.3$''$--0.5$''$ has been previously reported for
Arp~220, Markarian~273 and  IRAS 08572+3915,
and VV~114 (Soifer, et al. 1999, 2000, 2001).  In VV~114,
Arp~220,  and Markarian~273, two compact sources were seen at
12.5$\mu$m, while in IRAS 08572+3915 a single unresolved source was
detected that contains nearly all the bolometric luminosity of the
source.  No mid - infrared imaging of Markarian~463 has been
reported in the literature; as shown below, the present observations
show a compact core which emits essentially all the mid-infrared luminosity
of the galaxy.  Table 1 lists the basic properties of the
sources that were observed.

\section {Observations and Data Reduction}

The observations were made  using the low resolution spectrosopic mode
of the LWS (Jones \& Puetter, 1993) at the f/25 forward Cassegrain
focus of the  Keck~I 10-m Telescope. Briefly, LWS includes a grating
spectrometer which contains a  128$\times$128 pixel Si:As array.  The
spectral dispersion is 0.0375~$\mu$m pixel$^{-1}$, and the spatial
scale is 0.08~$\arcsec$~pixel$^{-1}$.  For the present spectroscopy, a
filter with bandpass from 8.1 to  13.0~$\mu$m was used for order
selection; 128 $\times$ 95  pixels (dispersion  $\times$
cross dispersion) were  un-vignetted, so a wavelength
coverage of 4.8~$\mu$m was available over a 7.6~$\arcsec$ slit length.

The present observations were obtained in May, September, and December
of 2000.  A log of the observations is presented in Table 2.  The
observations  were made at air masses between 1.1 and 1.5. The
atmospheric absorption in the observed spectra was monitored by
observing HR=BS~5340 in  May, 2000 and BS~8775 in September, 2000 and
December, 2000 at similar air masses.  For each source an image at
12.5~$\mu$m or 11.7~$\mu$m was obtained using the imaging mode of
the LWS before spectra were taken as part of the procedure of setting
up the target for spectroscopic observations.  Observations of stars
indicated the seeing had a  full width at half maximum (FWHM) between
0.3\arcsec and 0.4\arcsec. The   FWHM and the total on-source integration
times for the objects observed are listed in Table~2.

The observations used  chopping at a  frequency of $\sim$1.7~Hz and
nodding of the telescope every $\sim$30 seconds to suppress sky
emission. The telescope was nodded  in the same direction as the
chopping and the nodding amplitude was the same as the chopping
amplitude which is given in Table~2.  The chopping/nod direction was
set parallel to the slit so that, if the source structure allowed a
sufficiently small chopping amplitude, three dispersed images of the
objects were on the array. Observations were divided into runs of
$\sim$ 6 minutes each.  During the observations of the objects, the
telescope tracking was controlled using an offset guider with a
visible guide star.

For all these observations the slit was six pixels (0.48$''$)  wide
corresponding to a spectral resolution of 0.22$\mu$m or
$\lambda$/$\Delta\lambda$~$\sim$50 for a source uniformly filling  the
slit.  Because the seeing was typically less than the slit width,  the
spectral resolution was set by the image size, and not the slit width.
The seeing disk is close to the diffraction limit, which 
results in a variable image size; hence the
spectral resolution varies with wavelength.  In
general, the slit position angle on the sky was determined by the
location of an appropriately bright offset guide star. This limited
the observations to only one infrared component at a time in 
VV~114 and Markarian~273, even though there were two distinct
source components. In the case of    Arp~220 a guide star could be
used that allowed the slit to be oriented such that both nuclei,
separated by $\sim$1$''$ (Soifer et al. 1999),  could be observed
in a single spectrum.

The spectra were reduced by first combining the individual images to
subtract the sky contribution. The resultant images were then co-added
after appropriately shifting the images (a) parallel to the slit to
co-add the rows containing the object and (b) in the dispersion
direction to align the ozone emission present in the raw
un-sky-subtracted frames.  One-dimensional spectra were obtained by
extracting five rows (0.40$''$) centered on the maximum emission, and
subtracting from this the spectrum of the night sky extracted in the
two dimensional spectral image well away from the object.  In order to
increase the signal to noise ratios,  three pixels (half the slit
width) throughout the wavelength range  of the spectra were averaged
in a running average. Many of the spectra showed a deep absorption
band and consequently had locally extremely low signal to noise
ratios. For some of these spectra a running average of several more
pixels over the wavelength range of low signal was taken, reducing the
spectral resolution but increasing the signal to noise ratio. The
resultant spectral resolution is indicated in Table~2. The noise per
pixel at each wavelength was calculated from the ``sky'' well away from
the object.

Atmospheric and instrumental features, most notably the deep
atmospheric ozone absorption at 9.6~$\mu$m, were removed by dividing
the object spectrum by the spectrum of BS~5340 and multiplying by a
blackbody spectrum  at the star's effective temperature.  Since the
stellar image was generally smaller than the slit width, the star set
both the spectral resolution and wavelength  scale of the  spectrum,
and was often at a different wavelength than that of the associated
galaxy spectrum, as evidenced by sharp discontinuities at the edges of
the ozone absorption band. After extensive experimentation, it was
found better to use an average spectrum of 15 observations of BS~5340
than a single spectrum of a near-by (in time) stellar observation.

The wavelengths of the stellar spectrum were established by comparing
the observed stellar spectrum of BS~5340 with models of the earth's
transmission over this wavelength range  calculated by Lord
(1992). The final wavelength  solution for each object was refined,
within the bounds set by the ozone emission, by identifying different
astrophysical or terrestrial  atmosphere features in each object.  In
VV~114E$_{\rm{NE}}$, Markarian~273N, and Arp~220 it was possible to
use the emission of the PAH band at 11.25~$\mu$m (hereafter referred to 
as the 11.3$\mu$m feature) in the rest frame to
set the wavelengths. In Markarian~463E the continuum was strong enough
so that the atmospheric ozone absorption was detectable in individual
spectra and was used to establish the wavelength scale. For the
remaining three objects, an attempt was made to match the deep
atmospheric absorption at 9.6~$\mu$m before dividing by the stellar
spectrum.

\section {Results}

The mid-infrared spectra of the galaxy nuclei are presented in Figure~1.
The individual spectra are described in the following sections.

\subsection {VV~114}

The structure  of VV~114 is dominated  at infrared wavelengths by two
compact sources (Knop et al. 1994, Doyon et al. 1995, Soifer et al. 2001), 
both within the highly obscured eastern component of the source, 
referred to by Soifer et al.  as VV~114E$_{\rm{NE}}$ and
VV~114E$_{\rm{SW}}$. 2.2$\mu$m  imaging of VV~114E with NICMOS on HST
(Scoville et al. 2000) showed that VV~114E$_{\rm{NE}}$ is slightly
extended while VV~114E$_{\rm{SW}}$ is unresolved.  Mid-infrared
photometry of these sources (Soifer et al. 2001) reveals a deeper
depression in the mid-infrared SED in the brighter northeast component.

Figure~1 shows the spectra of VV~114E$_{\rm{NE}}$ and
VV~114E$_{\rm{SW}}$. The two peaks are separated by 1.8$''$ at
12.5$\mu$m (Soifer et al. 2001). With  seeing of 0.4$''$ the slit
width of 0.5$''$ ~isolates the separate peaks.  The spectra show that
indeed the 10$\mu$m depression of VV~114E$_{\rm{NE}}$ is significantly
deeper than in VV~114E$_{\rm{SW}}$. The  spectrum of
VV~114E$_{\rm{NE}}$ shows the characteristic feature of aromatics,
i.e. a peak  at 11.3$\mu$m, a shoulder at 8.6$\mu$m and the continued
rise to the shortest observed wavelengths.  The minimum flux in the
spectrum is at about 10.2 $\mu$m, consistent with the aromatic
spectrum of normal galaxies as seen by ISO (Helou et al.  2000).

By comparison the spectrum of VV~114E$_{\rm{SW}}$ shows a shallower
rise to longer wavelengths and a weak 11.3$\mu$m feature.   The 
equivalent width of the 11.3$\mu$m
feature  in VV~114E$_{\rm{SW}}$ is less than half that  in
VV~114E$_{\rm{NE}}$.  In addition, the minimum of the flux appears to
occur at shorter wavelengths, closer to 9.8$\mu$m in the rest frame of
the galaxy (although this might be due to a slight mismatch in the
correction for the terrestrial ozone absorption at 9.6$\mu$m.)

\subsection {IRAS~08572+3915}

IRAS~08572+3915 was found by Soifer et al (2000) to be a single very
compact source ($<$0.3$''$) at 12.5$\mu$m that contained nearly all
the mid-infrared luminosity detected in the $\sim$1$'$ IRAS beam
measurements. The mid-infrared spectrum of IRAS 08572+3915 is shown in
Figure~1.  The spectrum shows a strong depression
at 10$\mu$m, much greater than expected from the photometry reported
by Soifer et al., but consistent with that seen by Dudley and
Wynn-Williams (1997).  This is a result of the redshift of
this object, that puts the deep depression outside the sampling of the
photometric filters used by Soifer et al.

There is no evidence for PAH features in the spectrum. Rather the spectrum
is
characteristic of strong absorption by cold silicate dust overlying a
warmer background source (e.g.  Gillett et al. 1975b), showing a
deep minimum at 9.7$\mu$m.

\subsection {Markarian~273}

Soifer et al.(2000) showed that this galaxy is dominated by two
compact sources separated by 1$''$ at mid-infrared wavelengths.
Markarian~273NE is slightly extended, mostly in the east-west direction in
the NICMOS 2.2$\mu$m image of Scoville et al. (2000), while  Markarian~273SW
is unresolved in the same NICMOS 2.2$\mu$m image. Soifer et
al. showed  that the northern source is generally brighter in the
mid-infrared, but with a stronger depression at 10$\mu$m.

The mid-infrared  spectra of the separate nuclei of Markarian~273 are
shown in Figure~1. With seeing of
$\sim$0.3$''$ the 0.5$''$ slit adequately separates the spectra of the
two nuclei.  Markarian~273NE shows  a broad depression centered
shortward of 10 $\mu$m, rising steeply to shorter and longer
wavelengths. The spectrum shows a clear peak at 11.3$\mu$m and shoulder
at 8.6 $\mu$m  showing the presence of the aromatic PAH features.

The spectrum of Markarian~273SW is significantly different from that of
Markarian~273NE, i.e. much less  depression at 10 $\mu$m and
little evidence for an 11.3$\mu$m feature.  The rise in flux to
shorter wavelengths coincides with that seen in the northern source.
The flattening at $\lambda<$8.6$\mu$m is suggestive of PAH emission.

\subsection {Markarian~463E}

Markarian~463E  was found by Sanders et al. (1988a)  to be a high
luminosity galaxy with a ``warm'' mid-infrared color,
i.e. larger 25/60$\mu$m  flux density ratio than most ULIRGs in the IRAS
Bright Galaxy Sample (Soifer et al. 1987, Sanders et al. 1988b). The
optical spectroscopic classification of Markarian~463E is that of a
Seyfert galaxy (Miller \& Goodrich, 1990).

Imaging at 12.5~$\mu$m done in conjunction with the mid-infrared
spectroscopy shows Markarian~463E to be compact, with a source of
angular diameter $<$0.5$''$, or 500 pc for a redshift of z=0.05,
containing all the 12$\mu$m flux measured by IRAS.   The
mid-infrared spectrum is shown in Figure~1.  The spectrum shows a
drop from 8.0 to 9.6$\mu$m, a minimum at $\sim$10.0$\mu$m and then a
slightly rising continuum to longer wavelengths.  There is structure
in the spectrum at about 9.2$\mu$m which is most likely associated
with imperfectly corrected ozone absorption  in the
observations. There is a peak at 10.55 $\mu$m and no evidence
for peaks at  8.6 or 11.3 $\mu$m. The peak at 10.55$\mu$m could represent
[SIV] emission, which is strong in ISO-SWS spectra of comparable soures
(Sturm et al. 1999, Lutz et al. 2000).

\subsection {Arp~220}

Arp~220 is the closest ULIRG (Soifer et al. 1984, Sanders et al.1988b),
at a redshift z$=$0.018. The two nuclei seen at 12.5$\mu$m (Soifer
et al. 1999) are separated by 1$''$ or 360 pc.  The slit position angle
was 92\arcdeg~so that both nuclei were on the slit at the same
time. The seeing was $\sim$0.3$''$ for the observations, so that the
spectra of the nuclei  are well separated.

The spectra of the separate nuclei are shown in Figure~1.  The spectra
of both nuclei show strong minima at $\sim$10$\mu$m with roughly the
same depression compared to the flux density at 8$\mu$m.  The spectra of
the nuclei are quite similar, the fainter eastern component shows
somewhat stronger PAH emission, with a clear peak at 11.3 $\mu$m and a
shoulder at 8.6$\mu$m. In the brighter western source the PAH features
are  present but not as prominent.

\subsection {Diffuse PAH Emission}

The continuum sources are spatially compact,  consistent with the
sizes measured in previous imaging observations  (Soifer et al. 1999,
2000, 2001).  In addition to the compact continuum  structure, there
is extended emission in the 11.3$\mu$m PAH emission feature in some
galaxy nuclei in this sample.   In the three  sources where this feature
is prominent~--~VV~114E$_{\rm{NE}}$, Markarian~273NE and
Arp~220~--~the PAH emission is clearly extended.  This is revealed  in
the two-dimensional spectra of these sources shown in Figure~2.

Figure~2a displays a gray scale image of the two-dimensional spectrum
in the region of the 11.3$\mu$m PAH feature in Arp~220 with a contour
plot  of the intensity overlayed.  The continuum emission from the two
compact sources in Arp~220 is clearly seen along the spectral
dimension, while at 11.3$\mu$m a   band is seen bridging the 1.0$''$
(360 pc) gap between the two continuum sources.   There is weak 
PAH emission extending beyond the eastern source (see below).
In Figure~2b,  spatially extended PAH emission is seen extending to
one side of the continuum source in Markarian~273NE by about 0.7$''$
(500pc).

In VV114E$_{\rm{NE}}$ (Figure~2c)  there is clear 11.3$\mu$m 
emission extending in the spatial domain in both directons from the
continuum source. The PAH emission  extends $\sim$1.3$''$~(500 pc)
south and 0.7$''$~(300 pc) north of the continuum source.  An artifact
of the observations, which is most clearly seen in the contour
plot/grayscale of Figure~2c, is the apparent shift  of the wavelength
of the PAH feature along the spatial dimension.  The  shift is
probably a result of the seeing disk being small compared to the slit width
(0.5$''$), so that shifts in the centroid of the emission within the
slit translate in the spectra to apparent shifts in the wavelength
along the slit.

To show the spatial distribution of the PAH emission in Arp 220 in
more detail, we have constructed a PAH-only spectral image by
subtracting a modeled continuum emission.  The results are shown in
Figure 3.  
The spatial profile of the continuum was derived by summing the 2-D
spectral image (Figure 3a) over wavelengths longward of 12$\mu$m, where 
the continuum is strong and there is no strong emission line.  Then, 
the continuum
spectrum was constructed in two stages.  First, seven rows (i.e. pixels in
the spatial direction)  centered on the maximum emission of the western
nucleus were summed to create an average spectrum of the western source.
Second, the PAH contaminated spectral regions were masked and a 5th order
polynomial was fit to the average spectrum to produce a 1-D pure continuum
spectrum.  A 2-D continuum spectral image was then created by replicating
the spatial profile in the wavelength direction with an amplitude
determined by the 1-D continuum fit.

Figure 3b shows the continuum-subtracted PAH only spectral image.  It
clearly indicates the diffuse nature of the PAH emission.  It can also
be seen that the peak of the PAH emission is slightly displaced toward
the eastern nucleus.

Figure 3c compares the spatial distribution of the continuum and PAH
emission.  To produce these profiles, the flux over 0.187$\mu$m 
(five columns) centered on the
peak of the PAH emission was summed along the wavelength direction in
the PAH only image (Figure3b) and the modeled continuum image. 
This shows a striking difference between the 
two components.  Unlike the double-peaked continuum emission, the PAH
emission has a single peak $\sim$0.25'' east of the western
nucleus, and is nearly uniformly distributed from this peak
toward the eastern nucleus.  The PAH emission drops markedly at
the center of the western nucleus.  

\section {Discussion}

The spectra presented in Figure~1 show a variety of shapes and
features, but are quite similar to those seen previously in lower
spatial resolution observations (cf. Roche, et al. 1991, Dudley 1999).
The overall spectra are a combination of a smooth featureless
continuum, absorption due to intervening cold silicate material, and 
PAH emission features such as seen in the  nearby starburst galaxy
M82 (c.f. Gillett et al. 1975a, Willner et al. 1977, Roche et
al. 1991, Dudley, 1999, Sturm, et al. 2000).  In addition, the high
spatial resolution of the spectra, seen in Figure~2, resolves the
spectra of the individual nuclei and reveals the extended nature of the
PAH emission as compared to the compact continuum sources.  The
substantially higher spatial resolution (0.3--0.5$''$) of the  present
observations over previous observations permits the separation of the
spectra at different spatial locations.  In VV~114E$_{\rm{SW}}$ there
is a hint of extended  PAH emission in the two-dimensional spectrum
but excess pattern noise in the spectrum precludes determining 
with any confidence its contribution in the extracted spectrum.   In IRAS
08572+3915, Markarian~273SW and Markarian~463E  there is no evidence
for PAH emission in the spectrum of the compact source or in the
two-dimensional spectrum from which it was extracted.

\subsection {Spectra of The Compact Sources}

The spectra of Arp~220 E \& W, VV~114E$_{\rm{NE}}$ and
Markarian~273NE, corrected for the extended PAH emission, are shown
in Figure~4.  To correct for the extended PAH emission, the
strength of the PAH emission adjacent to the compact source was
measured in the two-dimensional spectrum, and a spectrum of M82,
normalized to this measured PAH strength, was subtracted from the
spectrum of the continuum  source of Figure~1.

A substantial uncertainty exists in the spectrum of the diffuse emission 
that is subtracted from the compact sources.
We {\it assume} that this spectrum is universal and like that of M82. Because
of the bright compact sources and the faint diffuse emission except in the
11.3$\mu$m PAH band, we could only measure the 11.3$\mu$m PAH strength off 
the nuclei.   
We could not independently determine the full spectrum off the nuclei.
While the smoothness of the resulting spectra as displayed in Figure 4
suggests the basic validity of our procedure, 
potential variation of PAH to continuum strengths (e.g. Tran et al. 2001)
lead to a substantial uncertainty.

Sturm et al. (2000) argue that the mid-infrared spectrum
of M82 suffers negligible extinction (A$_{V} < 5$ mag) based, among other 
arguments,  on model fitting
of its spectrum to the observed spectrum of the reflection nebula NGC 7023.
However, Willner et al (1977) estimated the extinction
to the infrared continuum in M82 to be A$_{V} \sim$ 25 mag from the measured
ratio of the Br$\alpha$ and Br$\gamma$
emission lines at 2.16 and 4.05 $\mu$m. If this latter determination of the
extinction applies to the PAH emitting region,
then presumably the M82 spectrum suffers a corresponding silicate
absorption of $\tau_{9.7\mu m}\sim 2$ in the observed spectrum.
The effect of larger extinction on
the 10$\mu$m spectrum of M82 is illustrated in Rigopoulou et al (1999).
Increasing the silicate absorption over that of M82 in the subtracted spectrum
would have the effect of significantly reducing the 9-11$\mu$m continuum
that is subtracted from the observed spectrum, compared to the 11.3$\mu$m
strength. In turn, this would decrease the inferred silicate absorption in
the continuum sources, by preserving a detectable continuum from 9-11$\mu$m.

In Figure~1, VV~114E$_{\rm{NE}}$ is the source with the strongest PAH
emission.  As seen in Figure~4, after the extended PAH emission is
subtracted from  the initial spectrum, little flux remains from 9 -- 11 $\mu$m.
There is a peak close to 11$\mu$m that may be residual
PAH emission. The residual continuum is consistent
with zero flux in the wavelength  range from 9 -- 11 $\mu$m,
suggesting that most of the apparent continuum over these wavelengths
in the spectrum of Figure~1  is due to an extended, overlying PAH spectrum.
The compact continuum source apparently suffers strong absorption 
by intervening cold silicate material.

In Markarian~273NE a significant PAH feature is seen in
Figure~1. Figure~4 shows a weak PAH shoulder feature remaining after
subtraction of the extended PAH emission suggesting that there is a 
residual PAH band associated with the compact source. This is consistent with
the observation of continuum detected in the subtracted spectrum over the 
entire 9-11$\mu$m range.

In Arp~220E the spectrum of Figure~1 shows a distinct PAH emission
peak at  11.3$\mu$m.  When the diffuse PAH emission is subtracted
from the observed spectrum, there is no evidence for residual PAH
emission in the compact source, and the continuum becomes consistent
with zero flux from 8.8 -- 11.5$\mu$m (Figure 4).  In Arp~220W, where the
spectrum of Figure~1 shows a shoulder of PAH emission on a  steeply
rising continuum, after subtraction of the diffuse PAH emission,  a
weak shoulder remains at 11.3$\mu$m, as does a measurable continuum 
in the 9--11 $\mu$m band.

The spectra of Figure~4 show that the underlying compact sources
suffer substantially greater foreground absorption at $\sim$10$\mu$m than 
would be
inferred from the spectra of Figure~1. Smith, Aitken and  Roche~(1991)
have attempted to correct the 10$\mu$m absorption apparent in the
spectrum of  Arp~220 for overlying PAH emission and derive   a peak
optical depth of $\tau_{9.7\mu m} >$5 which implies A$_{V}>$50 mag.
The spectra shown in Figure~4 suggest even greater absorption in the
depth  of the silicate band. From the spectra of Figure~4 we infer
peak optical depths $\tau_{9.7\mu m}$ (see below) in the range 8-15,  which
corresponds to optical extinctions A$_{V}$ of 80--150 mag.

\subsection {Detailed Comparison with ISO results for Arp 220}

Because it is the closest ULIRG, the results presented here for Arp 220 
warrant special discussion. A major finding from these observations is that 
the two nuclei in Arp 220 are not luminous sources of PAH emission.  As seen 
in Figures 
2 and 3, most of the PAH emission in Arp 220 originates in the off-nuclear
regions, and a significant fraction actually comes from the region
between the two nuclei.  Furthermore, the PAH emission drops
significantly near the center of the western nucleus, which dominates
the mid-infrared continuum luminosity.

The region producing the PAH emission is clearly distinct from the compact 
sources 
producing the bulk of the mid-infrared luminosity and most likely the
far-infrared luminosity as well. This argues that the
extinction toward the PAH emitting regions is not likely to be as
large as that inferred from the ISO data.  If
the extinction toward the PAH emitting region were as large as previously 
suggested (i.e. 50 magnitudes more than to the corresponding regions in M82),
star-forming regions directly associated with the PAH emitting regions would 
make a dominant contribution to the total continuum luminosity, which
is apparently not the case, as seen in Figures 2 and 3c. This is in contrast 
to the picture developed from the ISO data, which suggests that the strong
PAH emission seen in ULIRGs directly traces star formation that
dominates the luminosity output in these galaxies (Genzel et al. 1998,
Lutz et al. 1998, Rigopoulou et al. 1999, Tran et al, 2001).

The peak of the PAH emission, $\sim$0.25" east
of the western nucleus, corresponds to the peak of the warm molecular
hydrogen detected by Larkin et al. (1995).  This suggests that the
exciting source of the molecular hydrogen may also be responsible for
the PAH emission.

\subsection {The Compact Nuclei and the Diffuse PAH emission}

The excitation mechanism for PAH emission is generally believed to be
single UV/optical photon heating of aromatic molecules (see e.g. Sellgren, 
1984, Draine \& Anderson, 1985, Leger, D'Hendecourt and Defourneau, 1989).
The presence of substantial diffuse PAH emission associated
with the luminous compact sources shows that there is a significant
optical/UV photon intensity associated with the outer parts of the
sources.
The spectra of the underlying compact sources show large 10$\mu$m
optical depths. Based on the interstellar extinction  curve
(e.g. Mathis, 1990, Li \& Draine, 2001) this corresponds to an order
of magnitude greater extinction in the visible, and a factor of
$\sim$40 greater optical depth  in the UV.  

To roughly estimate the silicate
optical depths for the sources we assumed an underlying continuum that
is constant (in {\it f$_{\lambda}$}), and matched the apparent continuum 
with a silicate absorption profile as a function of wavelength from
Gillett et al. (1975b).  In the case of the most heavily absorbed
sources, where the continuum was not detected from 9-11$\mu$m, we
matched the observed continuum slope, corrected for PAH emission
in Figure 4, between 12 and 12.5$\mu$m (rest frame), and extrapolated the
inferred optical depth to the peak optical depth at 9.7$\mu$m.  
Because of the uncertainties in the shape
of the continuum and the simple technique employed for estimating 
$\tau_{9.7\mu m}$ 
these are crude estimates of the silicate optical depth.

The range of silicate optical depths and corresponding A$_{V}$ in these 
sources is quite
large. With  $\tau_{9.7\mu m}\sim$1 in Markarian~463E, A$_{V}$ is 
$\sim$10 mag, while in the most highly obscured sources like Arp~220E, the
silicate optical depths are as great as $\sim$15, corresponding to  
A$_{V}\sim$ 150 mag.
Given the much higher spatial resolution of the present observations,
it is not surprising that the extinction derived for the highly
embedded sources in Arp 220 are larger by a factor
of 2-3 than those determined from large beam ISO spectroscopic studies
(e.g. Sturm et al. 1996). The extinction to the continuum sources in Arp 220
is roughly an order of magnitude greater than that derived by Larkin et al.
(1995) from the  P$\beta$/Br$\gamma$ line ratios in these same nuclei. It is 
common to find increased
extinction determined at wavelengths that probe deeper into the regions; 
this suggests that we are still probing to $\tau\sim$1 at the
observed wavelengths.

The extinctions derived here are so
large that it is most unlikely that the UV/optical  photons that
excite the extended PAH emission arise internally in the compact
sources.  Much more likely is a large UV/optical radiation field
generated  due to hot, young stars in the periphery of the compact
sources.  The circumnuclear regions indeed have internal dust that
implies extinction of the ionizing starlight, but not in the huge
columns implied from the large silicate optical depths derived above.
The presence of such a circumnuclear environment is not
surprising, since powerful  circumnuclear starbursts have been
suggested for Arp~220 and Markarian~273 from  the distribution of CO
gas (Sakamoto et al. 1999, Downs \& Solomon, 1998).

We can crudely estimate the total luminosity associated with the
diffuse PAH emission by determining the total flux in the diffuse
11.3$\mu$m PAH  band and applying a bolometric correction based on an
assumed ratio of the PAH band to total luminosity. If we assume that
M82 has a typical PAH to bolometric luminosity ratio we derive a
bolometric correction  $\frac {L(PAH 11.3\mu m)} {L(bol)}=$0.0014 from
the spectrum   (Sturm et al. 2000) and  total far infrared luminosity
of M82 (Soifer et al. 1987). This correction factor is uncertain by
at least a factor of two, and depends critically on the amount
of additional extinction of the PAH spectrum (above that suffered in M82)
in the circumnuclear environments of these sources. To decrease
the ratio $\frac {L(PAH 11.3\mu m)} {L(bol)}$ by an order of magnitude from
that observed in M82 as a result of silicate absorption would require,
for the extinction curve of Li \& Draine (2001), an additional
overlying silicate optical depth at 9.7$\mu$m of 3.8, or A$_{V}=$45~mag 
above that affecting the M82 spectrum.

We have estimated the strength of the extended PAH emission (a)
coincident  with the continuum source and (b) integrated over the full
extent  of the PAH emission. We first measured the  strength of the
PAH band at a position adjacent to the continuum  sources of
Figure~2. To derive the PAH strength over the continuum  source, we
extrapolated this value to the spatial location of the continuum source.  
To  determine the total PAH emitting luminosity we integrated the flux in 
the PAH band
over its visible area in  the two-dimensional  spectra of
Figure~2. The PAH band strengths (as measured above) were scaled to the
measured fluxes of the continuum sources (matching bandpasses and 
spatial locations) from the imaging studies of these objects (Soifer et
al. 1999, 2000, 2001).  The uncertainty in converting PAH emission to
flux is at least 30\%, due to uncertainties in the location of the
object in the slit.

The results of these estimates for the luminosity corresponding  to
the PAH emission are given in Table 3 for Arp~220 E\&W, VV~114E$_{\rm{NE}}$
and Markarian~273N.  The luminosities attributed to PAH generating
sources, i.e. circumnuclear starbursts, lie in the range from a few
percent of the total bolometric luminosity in the Arp~220 sources, to
more than half of the total luminosity in the compact source in
VV~114E$_{\rm{NE}}$. 

If the bolometric correction that we have adopted
based on the PAH emission in M82 is an underestimate because of lower
extinction to the PAH emitting region in M82 than is appropriate for
these galaxies, the circumnuclear luminosities associated with extended
PAH emission in these systems would increase accordingly.  
In the case where the PAH to bolometric luminosity ratio is 
$\sim$10$^{-4}$, i.e. corresponding to A$_{V}\sim$50\,mag more than to
M82, the extended PAH emitting regions could account for 
roughly half the total bolometric luminosity of the sources. As discussed 
above for Arp 220, this seems extreme, since the continuum sources 
clearly dominate the mid-infrared flux in these observations and presumably
dominate the far-infrared continuum as well.

\subsection {Mid-Infrared Spectra As Diagnostics Of What Powers
Compact Sources?}

The spectra of the compact sources corrected for overlaying PAH
emission, show a wide range of silicate optical depths, from $\sim$1
to $\sim$15, and PAH strengths  ranging from no detectable PAH bands
to modest PAH bands.   Where the detectable silicate absorptions are
small, i.e. Markarian~463E, Markarian~273SW and VV~114E$_{\rm{SW}}$,
there is at most only weak evidence  for PAH emission and in two of
these cases strong evidence from optical spectroscopy for non-stellar
(AGN) ionizing sources dominating the luminosity (Koski, 1978, Miller
\& Goodrich 1990) .  In the other  sources where silicate absorption
is much greater, there is at best only weak PAH emission in the
mid-infrared specta.

Table 4 compiles the spectral classifications of these systems from a
variety of wavelength regions. The last column of Table 4,
from this work, is based on the strength of the 11.3$\mu$m PAH
band in the spectrum. The change in classification when correcting for
the overlying PAH emission shows the importance of utilizing the
highest possible spatial resolution.

Interpreting the mid-infrared spectra in terms of underlying sources
is difficult at best. The present observations demonstrate again that 
the central regions of ULIRGs
are quite complex. Continuum emitting regions obscured by many tens of 
magnitudes of extinction and less obscured diffuse
PAH emitting regions fall within the central few hundred parsecs of these 
sources. Resolving these components in the mid-infrared requires the high 
spatial resolution afforded by the largest groundbased telescopes.

In the majority of nuclei, the 10$\mu$m optical depth to the continuum
sources is very large, so it is difficult to infer the nature of the 
underlying power source.  The circumnuclear
luminosities that can be attributed to starbursts, from the diffuse
PAH emission, are substantial, and further highlights the picture of
heavily dust/gas enshrouded environments, powered by a highly obscured
source, with an overlying substantial starburst  region that
significantly affects the emergent spectra. The extent of the
circumnuclear starburst region, as determined from the spatial extent of
the PAH emission, is in the range 100 -- 500 pc, a size that is
resolvable at mid-infrared wavelengths  only in the closest of these
systems and only by the largest telescopes.   From the spectra presented
here, the large optical depths to the compact 10$\mu$m sources
suggest that even at these wavelengths the opacities are too large to probe
the true centers of these regions.  Whether the longer wavelength
diagnostics of SIRTF or HERSCHEL, with their comparatively poor spatial
resolution, can probe the cores of these systems is
unclear.

\section {Acknowledgments}

We thank J. Aycock and R. Campbell for assistance with the
observations and Barbara Jones, Rick Puetter and the  Keck team
that  brought the LWS into service, enabling these
observations. S. Lord kindly provided the  atmospheric model that was
used to establish the wavelength calibration.  G. Helou and L. Armus
kindly read a draft of this manuscript. We thank the anonomyous referee
for providing helpful comments.

The W.  M. Keck Observatory is operated as a scientific partnership
between the California Institute of Technology, the University of
California and the National Aeronautics and Space Administration. It
was made possible by the generous financial support of the W. M. Keck
Foundation.  We extend special thanks  to those of Hawaiian ancestry
on whose sacred mountain we are privileged to be guests.  Without
their generous hospitality, none of the observations presented herein
would have been possible.  This research has made use of the NASA/IPAC
Extragalactic Database which is operated by the Jet Propulsion
Laboratory, Caltech under contract with NASA.

B.T.S, K.M. and E.E. are  supported by grants from the NSF and
NASA. SIRTF is carried out at the Jet Propulsion Laboratory.  A.J.W
was supported by NASA grant NAG5-3042.

\newpage

\figcaption{} The mid-infrared spectra of the separate
nuclei plotted vs. rest wavelength. The logarithms of
the flux densities per wavelength interval
in relative units are plotted. Arrows indicate limits.
The spectrum of M82, binned to the
spectral resolution of the present data, is taken from
Sturm et al. (2000).

\figcaption{} Contour plots overlayed on gray-scale plots of the
two-dimensional spectra of Arp~220 (top), Markarian~273NE (middle)
and VV~114E$_{\rm{NE}}$ (bottom). In all cases, rest
wavelength is plotted along the abscissa
and spatial dimension is along the ordinate.  In each case the strong
continuum source is readily seen, while at the wavelength of the PAH
emission (11.25$\mu$m) extended emission along the spatial
dimension is also seen.  Slight wavelength shifts appear as a result
of the shifts of the centroid of the PAH band emission (in the x
spatial dimension) within the slit along the y spatial dimension.  In
Arp~220 the PAH emission is seen between the two continuum sources, in
VV~114E$_{\rm{NE}}$, PAH emission is seen to both sides of the continuum
source, while in  Markarian~273NE the PAH emission extends only to one side
of the continuum source.

\figcaption{} The spatial distribution of the 11.3$\mu$m PAH feature in 
Arp 220.  Panel a (top) is the observed two-dimensional spectrum, 
reproduced from the top panel of Figure 2. Panel b (middle) is the continuum 
subtracted two-dimensional spectrum that shows only the PAH band. The details
of how the continuum was removed are described in the text. Panel c 
(bottom) shows a one dimensional slice of the spatial distribution of the PAH
band (from Panel b) and the continuum.

\figcaption{}  The mid-infrared spectra of the nuclei,  plotted
vs. rest wavelength, and  corrected for the strength of the extended
PAH emission as displayed in Figure~2.  The correction was applied by
normalizing the spectrum of M82 (Sturm et al.  2000) to the strength
of the PAH band adjoining the continuum  source, and subtacting this
spectrum from the corresponding spectrum of Figure~1; see the text.
In each panel the top line is the spectrum from Figure 1, the bottom
line is the spectrum corrected for the overlying extended PAH emission.

\clearpage
\normalsize
\newpage

\normalsize


\begin{table}


\caption{ Basic Properties of  Observed Galaxies}

\smallskip
\begin{tabular}{l c c c c c }
\tableline\tableline
Name & z & log L & Spectrum &  linear scale \\
  &  &   L$_{bol}$[L$_{\odot}$] & OPT& Kpc/$''$ \\
\tableline

VV~114=IC 1623 &    0.020   & 11.62   & HII  &  400 \\

IRAS~08572+3915 & 0.058 & 12.10 &    LINER  & 1200 \\
Markarian~463E & 0.050 &12.03   &  Sey 1 & 1000\\
Markarian~273 & 0.037 & 12.13 & Sey  & 800\\
Arp~220 & 0.018 & 12.17 & LINER & 360 \\
\tableline

\end{tabular}

\end{table}

\clearpage

\begin{table}
\caption{Observing Log}
\begin{tabular}{c c c c c c c c}
\tableline\tableline
Object         &Date  & FWHM\tablenotemark{a}       & Obs\tablenotemark{b}
& Chop                        & P.A.\tablenotemark{c}
& \multicolumn{2}{c}{Averaging}\\
      &  &   & Time & Amp   & & Range \tablenotemark{d}    & Width
\tablenotemark{e}  \\
   & 2000 & \arcsec & s & \arcsec  & \arcdeg & $\mu$m  & $\mu$m \\
\tableline
 VV~114E$_{\rm{SW}}$ & Sep & 0.30 & 1200 &  5.0 &  69 & 8.91 --  9.64 & 0.26
\\
 VV~114E$_{\rm{NE}}$      & Dec & 0.35 &  400 & 10.0 & 166 & 8.98 -- 10.34 &
0.26    \\
 IRAS~08572+3915 & Dec & 0.42 &  200 & 10.0 & 157 & 9.37 --  9.72 & 0.26
\\
 Markarian~463E & May & 0.36 &  800 &  3.0 &  61 &  \nodata     & \nodata \\
 Markarian~273NE& May & 0.35 & 2600 &  3.0 &  83 & 8.98 -- 10.79 & 0.26
\\
 Markarian~273SW& May & 0.39 & 4800 &  3.0 &  83 &  \nodata     & \nodata \\
 Arp~220W       & May & 0.30 & 3000 &  3.0 &  92 & 9.11 -- 10.29 & 0.19
\\
 Arp~220E       & May & 0.30 & 3000 &  3.0 &  92 & 9.00 -- 10.11 & 0.34
\\
\tableline
\tablenotetext{a}{full width at half maximum of nearby PSF star}
\tablenotetext{b}{includes time of negative images on array; see text}
\tablenotetext{c}{position angle of slit measured east of north}
\tablenotetext{d}{wavelength range in rest frame over which additional
running
average greater than 0.11~$\mu$m was taken; see text}
\tablenotetext{e}{wavelength width of additional running average greater
than 0.11~$\mu$m; see text}
\end{tabular}
\end{table}

\begin{table}
\small

\caption{ Luminosities of PAH Emitting Regions}

\smallskip
\begin{tabular}{l  c c  c }
\tableline\tableline
Object & L$_{bol}$(cont)\tablenotemark{a} &
L$_{bol}$(PAH)\tablenotemark{b,d}
&
L$_{bol}$(PAH)\tablenotemark{c,d} \\
 &   & 10$^{10}$[L$_{\odot}$] & \\

\tableline

VV~114E$_{\rm{NE}}$& 5  & 2.5  & 12  \\

VV~114E$_{\rm{SW}}$ &  12  &  $<$1.2  & -- \\

IRAS~08572+3915&  150  & $<$15  & --\\

Markarian~273NE & 110   & 7  & 14 \\

Markarian~273SW & 25  &  $<$4  & --\\

Markarian~463E & 110  &  $<$11  & --\\
Arp~220W & 120  & 2  &6\tablenotemark{e}  \\

Arp~220E & 30   & 2  & 6\tablenotemark{e}  \\

\tablenotetext{a} {Continuum luminosities from Soifer et al. 1999,2000,
2001}
\tablenotetext{b} {Bolometric luminosities from diffuse PAH emission
overlaying compact source determined from
the strength of the extended PAH band and applying the bolometric correction
appropriate to M82 as described in the text. The uncertainty in this
bolometric correction is
at least a factor of 2}
\tablenotetext{c} {Total PAH-based bolometric luminosity from total
11.3$\mu$m
PAH emission as measured in two-dimensional spectrum of Figure~2.}
\tablenotetext{d} {The luminosity cited here should be considered a lower
limit because no correction has been applied for extinction to the PAH
region
beyond that suffered in the M82 spectrum}
\tablenotetext{e} {Total PAH luminosity is measured for the extended
region associated with both nuclei Arp~220E \& W}

\end{tabular}

\end{table}

\clearpage

\begin{table}
\small

\caption{ Spectral Classifications of Compact Nuclei }

\smallskip
\begin{tabular}{l  c c c c c  }
\tableline\tableline
Object &\multicolumn{5}{c} {  Spectral Class} \\

      &    Optical & Near IR & Mid-IR/ISO & X-ray &   Keck-Mid
IR\tablenotemark{a} \\

\tableline

VV~114E$_{\rm{NE}}$&HII\tablenotemark{1} &  SB\tablenotemark{4} &
AGN\tablenotemark{7} & -- & SB $\rightarrow$ ? \\

VV~114E$_{\rm{SW}}$ &HII\tablenotemark{1} & SB\tablenotemark{4} &
AGN\tablenotemark{7} & -- & SB?\\

IRAS~08572+3915& LINER\tablenotemark{1} & SB\tablenotemark{5}& --&
SB\tablenotemark{9} & AGN \\

Markarian 273NE & Sey2\tablenotemark{2} &  SB\tablenotemark{6} &
AGN\tablenotemark{8} & AGN\tablenotemark{10,12} & SB $\rightarrow$ ? \\

Markarian 273SW & Sey2\tablenotemark{2} &  SB\tablenotemark{6} &
AGN\tablenotemark{8} & AGN\tablenotemark{10,12} & AGN?\\

Markarian 463E & Sey1\tablenotemark{3} & -- & -- & AGN\tablenotemark{11} &
AGN\\
Arp~220E & LINER\tablenotemark{1} & SB\tablenotemark{6}& SB\tablenotemark{8}
& SB?\tablenotemark{10} & SB $\rightarrow$ ? \\

Arp~220W & LINER\tablenotemark{1} & SB\tablenotemark{6}& SB\tablenotemark{8}
& SB?\tablenotemark{10} & SB $\rightarrow$ ? \\

\tableline
\tablenotetext{a} {classifications are from strength of PAH band at
11.3$\mu$m, first entry
is from spectrum of Figure~1, second entry (where appropriate) is from
spectrum of Figure~3\\
SB= PAH feature clearly visible \\
SB? = PAH feature marginally visible \\
AGN = PAH feature not detected \\
\scriptsize
References: $^{1}$Veilleux et al. 1995, $^{2}$Koski 1978, $^{3}$Miller \&
Goodrich 1990, $^{4}$Doyon et al. 1995, $^{5}$Murphy et al. 2001,
$^{6}$Goldader et al. 1995, $^{7}$Laurent et al. 2000, $^{8}$Genzel et al.
1998, $^{9}$Risaliti et al.2000, $^{10}$Iwasawa 1999, $^{11}$ Ueno et al.
1996,$^{12}$Xia et al. 2001 \\
}
\end{tabular}

\end{table}
\clearpage

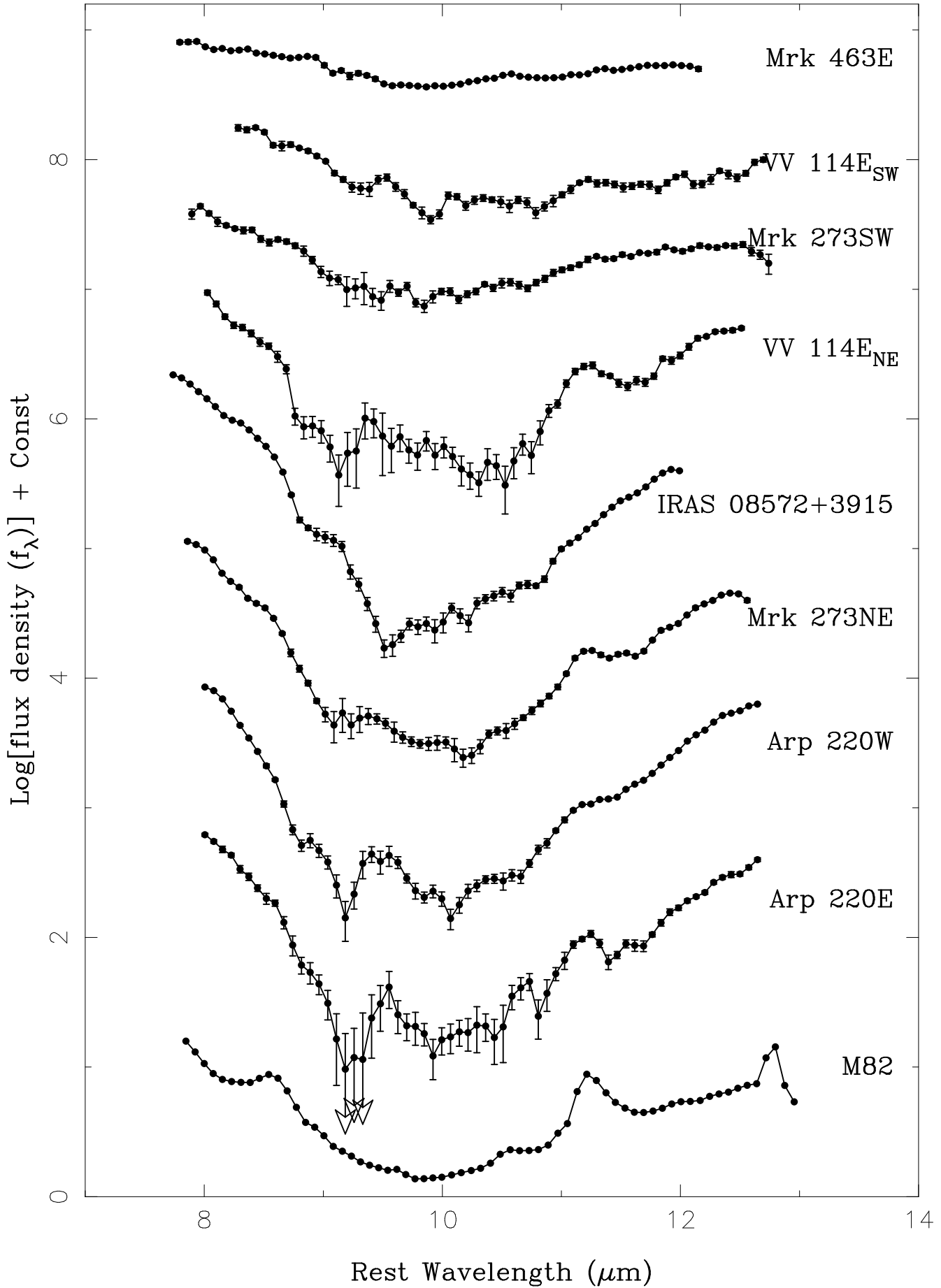
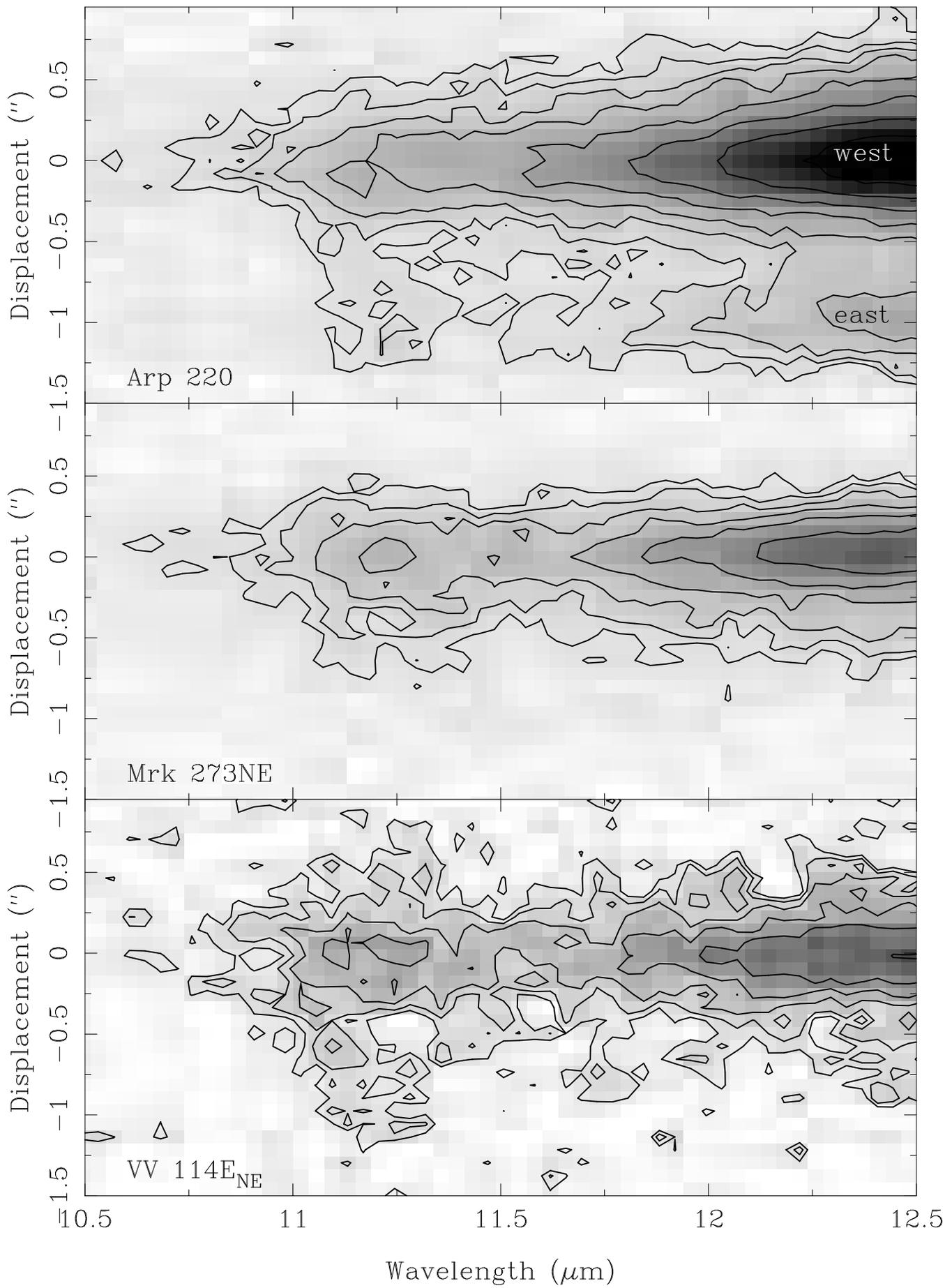
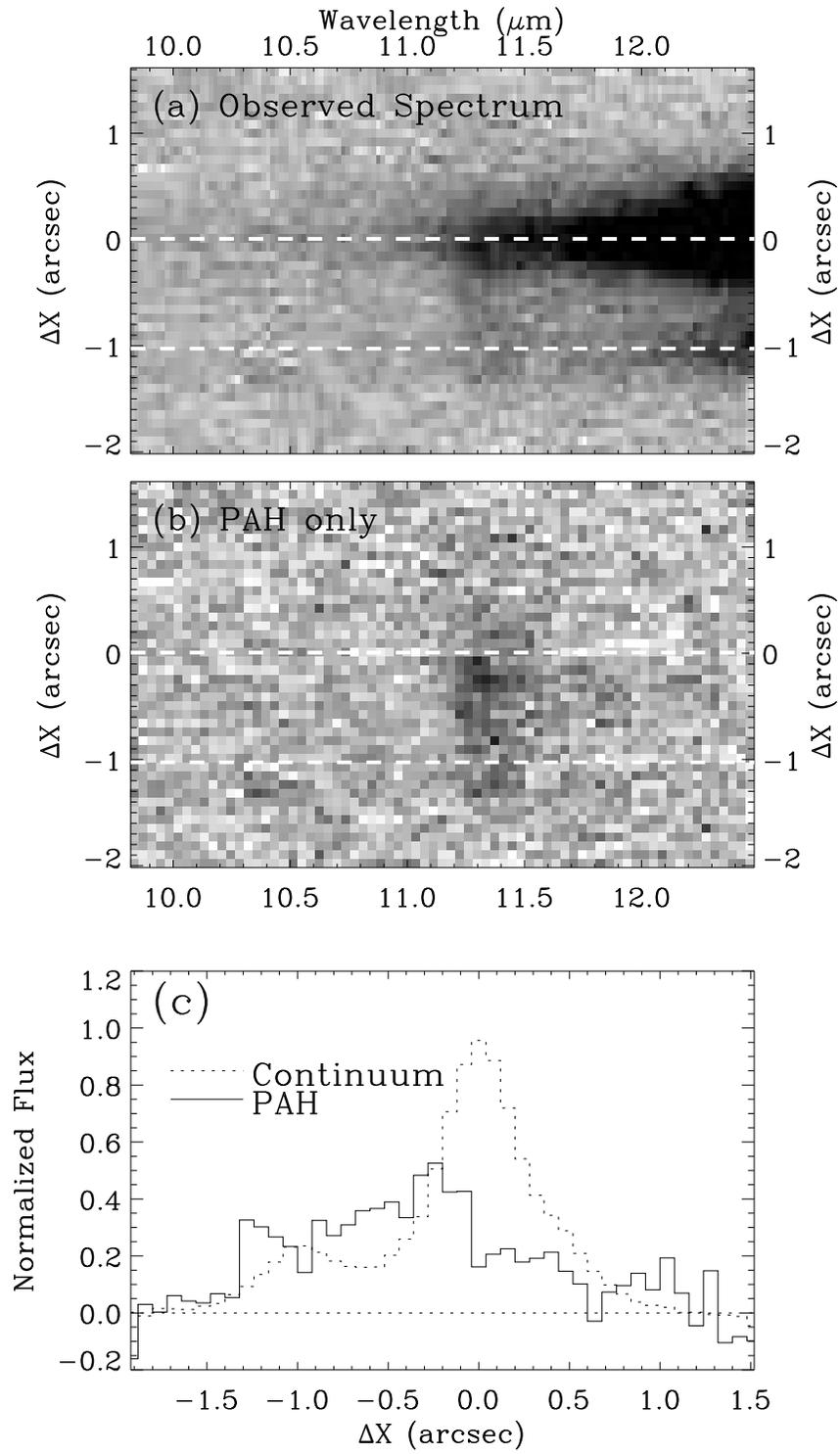
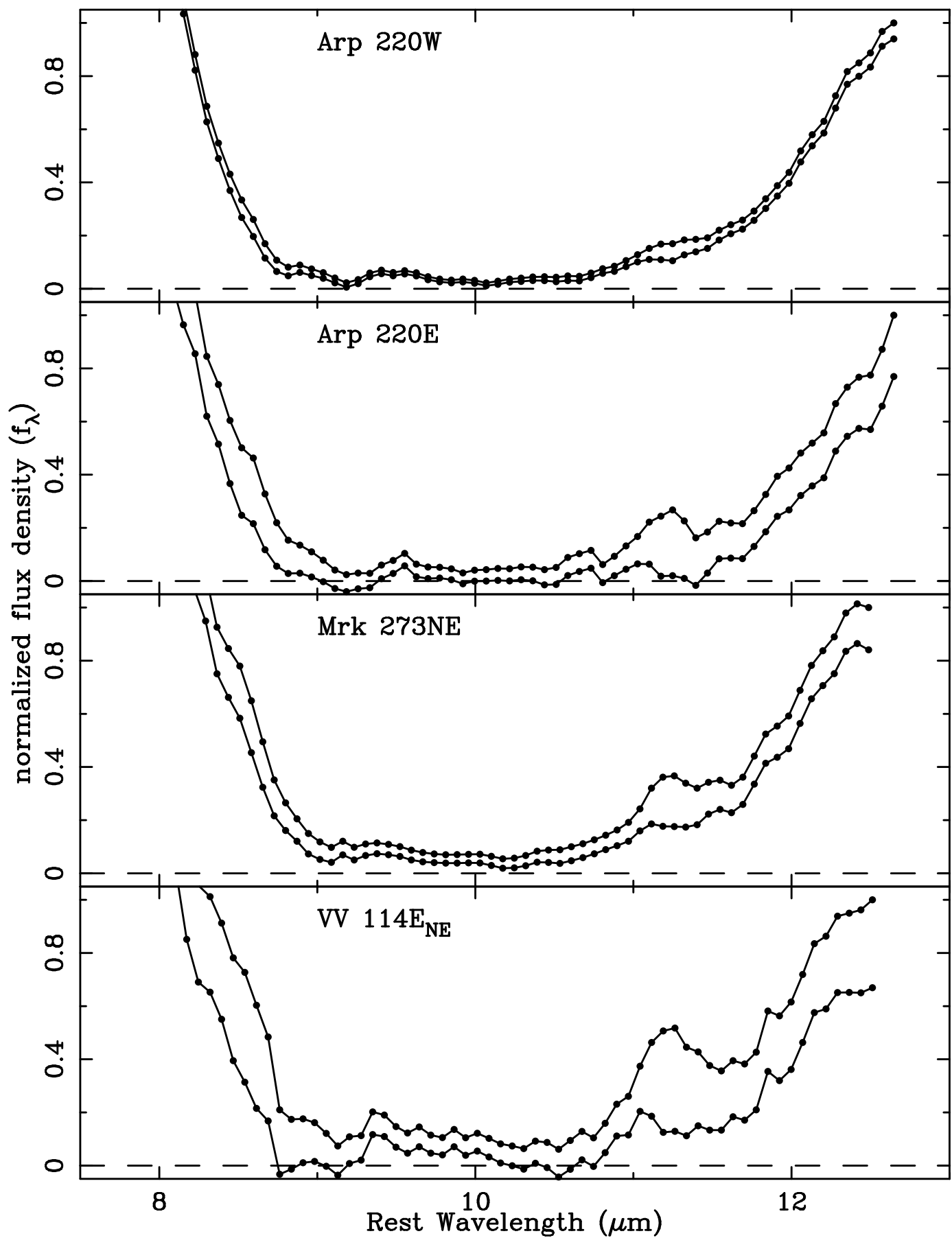
\end{document}